\documentclass{kluwer}    
\usepackage{psfig}
\newdisplay{guess}{Conjecture}

\begin{document}
\begin{article}
\begin{opening}
\title{Regular particle acceleration in relativistic jets
\thanks{{Partial
funding provided by RFBR grant 02-02-16900, INTAS grant 00491,
and Astronomy Programm "Nonstationary phenomena in
astrophysics"}}}
\author{Gennady \surname{Bisnovatyi-Kogan}}
\runningauthor{G.Bisnovatyi-Kogan} \runningtitle{Particle
acceleration in relativistic jets} \institute{Space Research
Institute RAN, Moscow, Russia
\\and\\Joint Institute of Nuclear Researches, Dubna, Russia}
\date{}

\begin{abstract}

Exact  solution is obtained for electromagnetic
field around a conducting cylinder of infinite length
and finite radius, with a
periodical axial current, when the wave length is
much larger than the radius of the cylinder.
The solution describes simultaneously the fields in
the near zone close to the cylinder, and transition to the wave
zone. Proper long-wave oscillations of such cylinder are studied.
The electromagnetic energy flux from the cylinder is calculated.
These solutions could be
applied for description of the electromagnetic field around
relativistic jets from active galactic nuclei and quasars
and particle acceleration inside jets.
\end{abstract}

\end{opening}

\section{Introduction}

 Objects of different scale and nature in the universe:
from young and very old stars to active galactic nuclei (AGN)
\cite{b},\cite{e},\cite{eh} show existence of
collimated outbursts - jets. Geometrical sizes of jets lay between
parsecs and megaparsecs. The origin of jets is not well understood
and only several qualitative mechanisms are proposed.
Theory of jets should answer to the question
of the origin of relativistic particles in outbursts
from AGN, where synchrotron emission is observed. Relativistic
particles, ejected from the central machine rapidly loose their
energy so the problem arises of particle acceleration inside the
jet, see reviews \cite{beg},\cite{bk93}.

It is convenient sometimes to investigate jets in a simple model
of infinitely long circular cylinder \cite{chf}. Magnetic fields
in collimated jets is determining its direction, and axial current
is stabilizing its elongated form at large distances from the source (AGN)
\cite{bkf}.
When observed with high angular resolution
these jets show structure with bright knots separated by
relatively dark regions \cite{b}, \cite{e},\ cite{tmw}.
 High percentages of polarization, sometimes
more then 50\%
in some objects, indicates the nonthermal nature of the
radiation which is well explained as the synchrotron radiation of
the relativistic electrons in a weak but ordered magnetic field.
Estimation of the life time of these electrons, based on the
observed luminosities and spectra, often gives  values much less
the their kinematic ages $t_k/c $, where $d$ is the distance of
the emitting point from the central source. Because the jet flow
most likely originated from an outburst or continuous outflow from
the central source, there is a necessity of continuous
reacceleration of electrons
in the jets in order to explain the observations.
Acceleration mechanism for electrons in extragalactic jets, proposed
in \cite{bkl}
considers that intense long-wavelength electromagnetic
oscillations accompany a relativistic jet as a result of the
non-steady mechanism of the jet's generation in the nucleus of the
source. The electromagnetic wave amplitudes envisioned are
sufficient to give in situ acceleration of electrons to the very
high energies observed $> 10^{13} $ eV.
It was assumed that jets are formed by a sequence of outbursts from the nucleus with
considerable charge separation at the moment of the outburst \cite{bkf}.
The direction of motion the outburst is determined
by the large-scale magnetic field. The outburst are accompanied by
an intense electromagnetic disturbance which propagates outward
moving with the jet material in the direction of the large scale magnetic
field. It was suggested in \cite{bkf} that a toroidal magnetic field,
generated during the outbursts is important for the lateral confinement of
of the jet.

When the plasma density in surrounding medium is small, the
electromagnetic wave generated by the nonpotential plasma oscillations of the
confined body is emitted outside and can accelerate particles.
When the emitted wave is strong enough it washed out the medium
around and the density can become very small, consisting only of the
accelerated particles. The action of the oscillating knot is
similar to the action of the pulsar, as inclined magnetic rotator. Both emit
strong electromagnetic waves, which could effectively accelerate particles
\cite{pa}, \cite{go}.
Long-periodic proper oscillations in the plasma cylinder with a finite radius,
and emission of electromagnetic waves had been studied in \cite{bkl},
and in a simpler model in \cite{bk}. Enhanced oscillations in such cylinder
have been studied in \cite{bk04}. Both models are represented below.

\section{Cylinder with oscillating current}

Consider infinitely conducting circular cylinder in vacuum.
This model is valid at low density in surrounding plasma, which cannot
screen the emitting electromagnetic wave. Maxwell equations are \cite{ll}

\begin{equation}
\label{maxh}
{\rm div}\,{\bf B}=0,
\quad
{\rm rot}\,{\bf B}=\frac{1}{c} \frac{\partial {\bf E}}{\partial t}
 + \frac{4\pi}{c} {\bf j},
\end{equation}

\begin{equation}
\label{maxe}
{\rm rot}\,{\bf E}=-\frac{1}{c} \frac{\partial {\bf B}}{\partial t},
\quad
{\rm div}\,{\bf E}=4\pi \rho_e.
\end{equation}
For periodic oscillations with all values $\sim \exp({-i\omega t})$
they read

\begin{equation}
\label{maxth}
{\rm div}\,{\bf B}=0, \,\,\, {\rm rot}\,{\bf B}=
-\frac{i\omega}{c}{\bf E}+\frac{4\pi}{c} {\bf j},\,\,\,
{\rm rot}\,{\bf E}=\frac{i\omega}{c}{\bf B},\,\,\,
{\rm div}\,{\bf E}=0.
\end{equation}
We use the same definitions for all complex values depending on
coordinates. Consider infinitely long cylinder with zero charge density,
where in the cylinder coordinate system
$(r, \phi, z)$ the only nonzero components are
$E_z,\,\, B_{\phi},\,\, j_z$, and
$\partial/\partial\phi=\partial/\partial z=0$.
Only two valid equations remain from the system
(\ref{maxth}):

\begin{equation}
\label{ref5}
\frac{dE_z}{dr}+\frac{i\omega}{c}B_{\phi}=0,\quad
\frac{1}{r}\frac{d(rB_{\phi})}{dr}+\frac{i\omega}{c}E_z-\frac{4\pi}{c}j_z=0;
\end{equation}
from which, we obtain equation for
$E_z$:

\begin{equation}
\label{ref7} \frac{1}{r}\frac{d}{dr}\left(r\frac{d E_z}{d
r}\right)+\frac{\omega^2}{c^2}E_z+\frac{4\pi i \omega}{c^2} j_z=0.
\end{equation}

\section{Vacuum solution}

In vacuum $j_z=0$. Using non-dimensional variable
$x=r\omega/c$, we obtain from (\ref{ref5}),(\ref{ref7})

\begin{equation}
\label{ref8} x^2 E_z''+ xE_z' + x^2 E_z=0, \qquad B_{\phi}=iE_z'.
\end{equation}
Here $'$ \,\, denote differentiation over $x$. The equation
(\ref{ref8}) belongs to Bessel type and has a solution

\begin{equation}
\label{ref9} E_z=C_1J_0(x)+C_2 Y_0(x), \qquad
B_{\phi}=-i[C_1J_1(x)+C_2 Y_1(x)].
\end{equation}
Relations for Bessel functions are
\cite{gr}

\begin{equation}
\label{ref10} J_0'(x)=-J_1(x), \qquad Y_0'(x)=-Y_1(x).
\end{equation}
General solution for physical values, with account of the time
dependence, is obtained from the real part of the complex solution
at

\begin{equation}
\label{ref11} \exp{(-i\omega t)}=\cos \omega t-i\sin \omega t,
\qquad C_1=C_1^{(r)}+iC_1^{(i)},\quad C_2=C_2^{(r)}+iC_i^{(i)}.
\end{equation}
The general solution in vacuum is

\begin{equation}
\label{ref12} E_z=[C_1^{(r)}J_0(x)+C_2^{(r)} Y_0(x)]\cos \omega t
+ [C_1^{(i)}J_0(x)+C_2^{(i)} Y_0(x)]\sin \omega t,
\end{equation}

\begin{equation}
\label{ref13} B_{\phi}=-[C_1^{(r)}J_1(x)+C_2^{(r)} Y_1(x)]\sin
\omega t + [C_1^{(i)}J_1(x)+C_2^{(i)} Y_1(x)]\cos \omega t.
\end{equation}
The boundary condition far from the cylinder follows from the
demand that there exist only expanding wave.
It means, that only functions depending on the combination
$(x-\omega t)$ survive. Using asymptotic of Bessel functions at large arguments
\cite{gr}

$$J_0(x) \approx \sqrt{\frac{2}{\pi x}}\cos (x-\frac{\pi}{4}), \quad
J_1(x) \approx \sqrt{\frac{2}{\pi x}}\sin (x-\frac{\pi}{4}), $$
\begin{equation}
\label{ref14}
Y_0(x) \approx \sqrt{\frac{2}{\pi x}}\sin
(x-\frac{\pi}{4}), \quad Y_1(x) \approx -\sqrt{\frac{2}{\pi
x}}\cos (x-\frac{\pi}{4}), \quad {\rm at} \quad x \gg 1.
\end{equation}
we obtain for the expanding wave
 $C_1^{(i)}= -C_2^{(r)}$, $C_2^{(i)}= C_1^{(r)}$, leading to the
following solution at large distances

\begin{equation}
\label{ref15} E_z \approx \sqrt{\frac{2}{\pi x}} [C_1^{(r)}\cos
(x-\frac{\pi}{4}-\omega t)+C_2^{(r)}\sin (x-\frac{\pi}{4}-\omega
t)], \qquad B_{\phi}=-E_z.
\end{equation}
The general vacuum solution, satisfying conditions at infinity
reads as

\begin{equation}
\label{ref16}
E_z=[C_1^{(r)}J_0(x)+C_2^{(r)} Y_0(x)]\cos \omega t
+ [-C_2^{(r)}J_0(x)+C_1^{(r)} Y_0(x)]\sin \omega t,
\end{equation}
\begin{equation}
\label{ref17}
B_{\phi}=-[C_1^{(r)}J_1(x)+C_2^{(r)} Y_1(x)]\sin
\omega t + [-C_2^{(r)}J_1(x)+C_1^{(r)} Y_1(x)]\cos \omega t.
\end{equation}

\section{Solution inside the cylinder}

The equation in the matter are

\begin{equation}
\label{ref18} x^2 E_z''+ xE_z' + x^2 E_z+\frac{4\pi i}{ \omega}
x^2 j_z=0, \qquad B_{\phi}=iE_z'.
\end{equation}
A solution of the non-uniform linear equation
(\ref{ref18}) is a sum of a general solution of the uniform
equation, and a particular solution of the non-uniform
one  ${\cal E}_0(x)$.

\begin{equation}
\label{ref19} E_z={\cal E}_1J_0(x)+{\cal E}_2 Y_0(x)+{\cal
E}_0(x).
\end{equation}
The function $Y_0(x)$ is singular at $x=0$, so for a finite solution
${\cal E}_2=0$.
We look for a particular
solution in the form $E_z={\cal E}(x)J_0(x)$.
First order equation with respect to $F={\cal E}'$
follows from (\ref{ref18})

\begin{equation}
\label{ref20} x^2(F'J_0 + 2F J_0') + x F J_0+\frac{4\pi i}{
\omega} x^2 j_z=0.
\end{equation}
From this equation we obtain the general solution for
the amplitude of the electric field in the matter, in presence
of periodic EEF:

\begin{equation}
\label{ref21} E_z=-\frac{4\pi i}{ \omega}
J_0(x)\int_0^x\frac{dy}{yJ_0^2(y)}\int_0^y zJ_0(z)j_z(z)dz + {\cal
E}_1J_0(x).
\end{equation}
Consider waves much longer than the radius of the cylinder
$r_0$

\begin{equation}
\label{ref22} x_0=\frac{\omega r_0}{c} \ll 1.
\end{equation}
Than use expansion
  at $x\ll 1$ \cite{gr},

\begin{equation}
\label{ref23} J_0 \approx 1-\frac{x^2}{4}, \quad J_1 \approx
\frac{x}{2}, \quad Y_0 \approx \frac{2}{\pi}\ln{\frac{x}{2}},\quad
Y_1 \approx -\frac{2}{\pi x}.
\end{equation}
Using (\ref{ref23}) we obtain from
 (\ref{ref21}) the solution for long waves

\begin{equation}
\label{ref24} E_z=-\frac{2 i\omega}{c^2} \int_0^x
I_z(y)\frac{dy}{y} + {\cal E}_1,
\end{equation}
where $I_z(r)\equiv I_z(y)$ is the complex amplitude of the electrical
current inside a cylindrical radius $r=cy/\omega$

\begin{equation}
\label{ref25} I_z =2\pi \int_0^rj_z r\, dr
=2\pi\frac{c^2}{\omega^2}\int_0^xj_z x\,dx
\end{equation}
Complex values: the function $I_z(r)$ and the constant
 ${\cal E}_1$ are

\begin{equation}
\label{ref26}
I_z = I_z^{(r)}+i I_z^{(i)}, \qquad {\cal E}_1={\cal
E}_1^{(r)}+i{\cal E}_1^{(i)}.
\end{equation}

\section{Matching of solutions and long-wave limit}

The total electrical current through the cylinder
$I_0 = I_z(r_0)$, and fields on its surface (inside)
  $E_0=E_z(r_0)$, $B_0=B_{\phi}(r_0)$
(real parts of complex relations) are defined as

$$
 I_0 = I_0^{(r)}\cos\omega t+ I_0^{(i)}\sin \omega t,\quad E_0=
$$
\begin{equation}
\label{ref27}
=\biggl[\frac{2 \omega}{c^2}\int_0^{x_0}
I_z^{(i)}(y)\frac{dy}{y}+ {\cal E}_1^{(r)}\biggr]\cos \omega t +
\biggl[-\frac{2 \omega}{c^2}\int_0^{x_0}I_z^{(r)}(y)\frac{dy}{y}
+{\cal E}_1^{(i)}\biggr]\sin\omega t,
\end{equation}
$$
B_0=\biggl[\frac{2 \omega}{c^2}
\frac{I_0^{(r)}}{x_0}+ \frac{x_0}{2}{\cal E}_1^{(i)}\biggr]\cos \omega t +
\biggl[\frac{2\omega}{c^2}\frac{I_0^{(i)}}{x_0}
-\frac{x_0}{2}{\cal E}_1^{(r)}\biggr]\sin\omega t.
$$
At small $x_0$ the external solution on the surface of the cylinder
is written as
\begin{equation}
\label{ref28}
E_{z0}
=[C_1^{(r)}+C_2^{(r)}\frac{2}{\pi}\ln\frac{x_0}{2}]\cos \omega t
+ [-C_2^{(r)}+C_1^{(r)}\frac{2}{\pi}\ln\frac{x_0}{2}]\sin \omega t,
\end{equation}
$$B_{\phi 0}=-[C_1^{(r)}\frac{x_0}{2}-C_2^{(r)}\frac{2}{\pi x_0}]\sin
\omega t + [-C_2^{(r)}\frac{x_0}{2}-C_1^{(r)}\frac{2}{\pi x_0}]\cos \omega t.
$$
All field components are continuous at the cylinder surface in absence
of the surface charges and currents. Matching magnetic and electrical
fields we obtain using
(\ref{ref27}), (\ref{ref28})
 coefficients in the solution of the
external electromagnetic field, which are determined by the periodic
electrical current in the cylinder:

\begin{equation}
\label{ref32}
C_1^{(r)}\left(\frac{2}{\pi x_0}+\frac{x_0}{\pi}\ln\frac{x_0}{2}\right)=
-\frac{2 \omega}{c^2}\frac{I_0^{(r)}}{x_0}
-\frac{x_0 \omega}{c^2}\int_0^{x_0}I_z^{(r)}(y)\frac{dy}{y},
\end{equation}
\begin{equation}
\label{ref33}
C_2^{(r)}\left(\frac{2}{\pi x_0}+\frac{x_0}{\pi}\ln\frac{x_0}{2}\right)=
\frac{2 \omega}{c^2}\frac{I_0^{(i)}}{x_0}
+\frac{x_0 \omega}{c^2}\int_0^{x_0}I_z^{(i)}(y)\frac{dy}{y}.
\end{equation}
\begin{equation}
\label{ref34}
 {\cal E}_1^{(r)}=
C_1^{(r)}+C_2^{(r)}\frac{2}{\pi}\ln\frac{x_0}{2}
-\frac{2 \omega}{c^2}\int_0^{x_0}
I_z^{(i)}(y)\frac{dy}{y},
\end{equation}
\begin{equation}
\label{ref35}
{\cal E}_1^{(i)}=-C_2^{(r)}+C_1^{(r)}\frac{2}{\pi}\ln\frac{x_0}{2}
+\frac{2 \omega}{c^2}\int_0^{x_0}I_z^{(r)}(y)\frac{dy}{y}.
\end{equation}
Consider a case when the resulting electrical current produced by external
EEF is purely sinusoidal, $I_z^{(r)}=0$.
Than it follows from (\ref{ref32}),(\ref{ref35})

\begin{equation}
\label{ref36}
C_1^{(r)}=0, \quad {\cal E}_1^{(i)}=-C_2^{(r)}.
\end{equation}
At $x_0 \ll 1$ we neglect logarithmic terms in (\ref{ref32})-(\ref{ref33}),
and terms with integrals in (\ref{ref32})-(\ref{ref35}), so the details of the
current distribution over the cylinder radius are of a little importance.
In this approximation

\begin{equation}
\label{ref37}
C_2^{(r)}=\frac{\pi \omega}{c^2}I_0^{(i)}, \quad
{\cal E}_1^{(r)}=\frac{2 \omega}{c^2}\ln\frac{x_0}{2}I_0^{(i)}.
\end{equation}
The solution for the electromagnetic field of the
long wave emitted by the cylinder
with the sinusoidal electric current, starting from the surface
of the cylinder until the wave zone, follows from
(\ref{ref16}):

$$
E_z= \frac{\pi \omega}{c^2}I_0^{(i)}[Y_0(x)\cos \omega t
-J_0(x)\sin \omega t],\quad
$$
\begin{equation}
\label{ref38}
B_{\phi}=-\frac{\pi \omega}{c^2}I_0^{(i)}[Y_1(x)\sin
\omega t +J_1(x)\cos \omega t].
\end{equation}
Near the cylinder we have, with account of the expansions
(\ref{ref23})

\begin{equation}
\label{ref39}
E_z= \frac{\pi \omega}{c^2}I_0^{(i)}(\frac{2}{\pi}\ln\frac{x}{2}\cos \omega t
-\sin \omega t)
\end{equation}
$$\approx \frac{2\omega}{c^2}I_0^{(i)}\ln\frac{x}{2}\cos \omega t=
\frac{2\omega}{c^2}I_0^{(i)}\ln\frac{r\omega}{2c}\cos \omega t ,
$$
\begin{equation}
\label{ref40}
B_{\phi}=-\frac{\pi \omega}{c^2}I_0^{(i)}(-\frac{2}{\pi x}\sin
\omega t +\frac{x}{2}\cos \omega t)\approx \frac{2 \omega}{c^2 x}I_0^{(i)}\sin
\omega t=\frac{2I_0^{(i)}}{cr}\sin\omega t.
\end{equation}
Note, that in the near zone of the long wave the magnetic field adiabatically follows the
current through the cylinder \cite{ll}. The distribution of the electrical field
is similar to the one of the linearly growing current in the cylinder  \cite{bkmn}.
At large $r$ we obtain from (\ref{ref38}),
(\ref{ref14}) the expanding cylindrical
wave

\begin{equation}
\label{ref41}
B_{\phi}=-E_z
=-\frac{\pi \omega}{c^2}\sqrt{\frac{2}{\pi x}}I_0^{(i)}\sin (x-\frac{\pi}{4}-\omega t)
\end{equation}
$$
=-\frac{1}{c}\sqrt{\frac{2\pi\omega}{c r}}I_0^{(i)}\sin [\frac{\omega}{c}(r-ct)-\frac{\pi}{4}].
$$

\section{Electromagnetic energy flux from jet}

\begin{figure}
\centerline { \psfig{figure=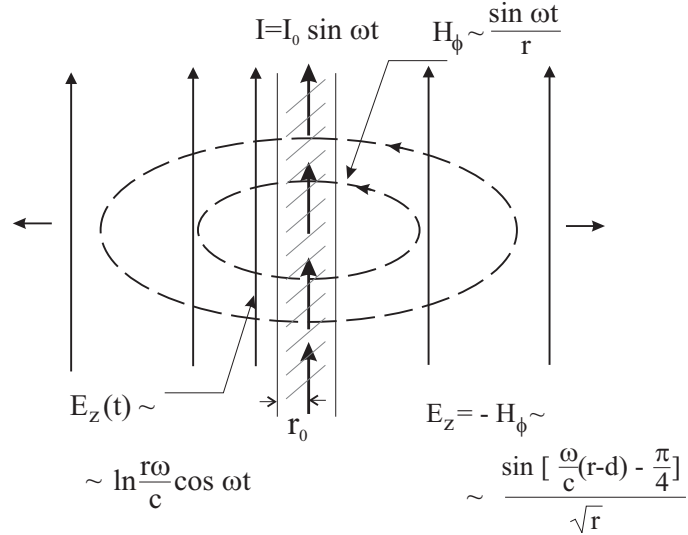,width=9cm}}
\caption{Magnetic and electrical fields around the infinite cylinder
with the radius
$r_0$, and low-frequency $\omega \ll c/r_0$,
sinusoidal electrical current along the cylinder axis.
In the near zone electrical and magnetic fields are varying in antiphase,
and far from the cylinder
 $r \gg c/\omega$ the expanding cylindrical electromagnetic wave is formed, with
 $E_z=-B_{\phi}$.}
\label{fig1}
\end{figure}
 Strong electromagnetic wave generated by oscillations may accelerate
effectively particles at large distances from the nucleus near the jet, as well as
at larger radiuses \cite{bkl}. Let us estimate the energy flux in the electromagnetic
wave radiated by the jet of the length
$l$, and radius $r_0$. If $n_e$ is the electron density producing the
electrical current, with  $e$ as electron charge, than, with account
of (\ref{ref41}) the Pointing flux
$P=\frac{c}{4\pi}[{\bf EB}]$ through the cylinder surface is

\begin{equation}
\label{ref42}
F=2\pi r_0 l P=\frac{\pi l\omega}{2c^2} I_0^2.
\end{equation}
For the amplitude of the electrical current along the cylinder radius
$I_0= \pi r_0^2 n_e ce$, we obtain the energy flux
from the jet in the form:

\begin{equation}
\label{ref43}
F=\frac{\pi^3}{2} e^2lr_0^4 \omega n_e^2
\approx 2\cdot 10^{49}
\,\mbox{erg/s}\,\frac{l}{1\,\mbox{kpc}}\left(\frac{r_0}{1\,\mbox{pc}}\right)^4
\frac{T}{100\,\mbox{yr}}\left(\frac{n_e}{10^{-10}\,\mbox{cm}^{-3}}\right)^2.
\end{equation}
Here $T=\frac{2\pi}{\omega}$ is the period of the electromagnetic wave.
Part of the radiated energy is used for particle acceleration
up to very large energies \cite{bkl}, and support the jet radiation at
different energy regions of electromagnetic spectrum.

\section{Generation of strong electromagnetic wave by proper
oscillations in a separate blob.}

 The mechanism of shock acceleration of particles, often considered \cite{eh},
 is not certain, it is unlikely that shock
 acceleration can give a fairly uniform brightness jet as observed in
 some cases. The mechanisms of magnetic field
 reconnection \cite{rl}
 and plasma turbulence acceleration \cite{eh}
 are also highly uncertain.
When plasma density in the surrounding medium is small, the
electromagnetic wave generated by the nonpotential plasma oscillations of the
confined body is emitted outside and can accelerate particles.
When the emitted wave is strong enough it washed out the medium
around and the density may become very small, consisting only of the
accelerated particles.
The solution of the whole problem of the dynamical behavior
of the confined knots embedded into the large scale elongated magnetic
field and producing the toroidal field can be solved by self-consistent
calculations of the knot oscillations, using together the hydrodynamical
and complete Maxwell equations.
In order to estimate properties of a long wave radiation of by
oscillating knot we solve instead the idealized problem having the analytical
solution.
Consider linear plasma oscillations of the infinitely long uniform
cylinder. The problems of such kind have been
intensively studied for plasma wave-guides \cite{kon}.
The main difference in this
problem is another boundary conditions which suggest vacuum state around the
cylinder. When considering linear electromagnetic oscillations in the static
plasma cylinder, only Maxwell equations (\ref{maxth})
with time dependence in the form $ \sim \, \exp (-i \omega t)$
are needed.
 The
background constant field $ B_z=B_0$ is adopted. Dependence if $ j$ on
${\bf E,\, B \;}$ and $ B_0 $ can be obtained, using the expression for
the dielectric permeability

\begin{equation}
\label{ref45}
\epsilon _{ij}\;=\; \delta_{ij} \,+\, {4 \pi i \over \omega}
\,[\sigma _{ij}(e)\,+\,\sigma_{ij}(p)],\,\,\,
j_i\,=\,[\sigma _{ij}(e)\,+\,\sigma_{ij}(p)]\,E_j
\end{equation}
Here we consider for simplicity pure hydrogen plasma. The components for
$ \sigma_{ij}(\alpha) $ in the case of perfect conductivity are
\cite{kon}

\begin{equation}
\label{ref46}
\sigma_{11}(\alpha)\,=\,\sigma_{22}(\alpha)\,=\,{i \over 4 \pi}
{\omega_{p\alpha}^2\, \omega \over \omega^2\,-\,\omega_{B\alpha}^2},
\end{equation}
$$\sigma_{12}(\alpha)\,=\,-\sigma_{21}(\alpha)\,=\,-{1 \over 4 \pi}
{\omega_{p\alpha}^2\, \omega_{B\alpha} \over \omega^2\,-\,\omega_{B\alpha}^2},\quad
 \sigma_{33}(\alpha)=
\,{i \over 4 \pi}{\omega_{p\alpha}^2 \over \omega}  $$
where $\omega_{p\alpha}$ and $\omega_{B\alpha}$ are plasma and Larmor
frequencies of electrons (e) and protons (p):
$ \omega_{p\alpha}={4 \pi n_0 e_{\alpha}^2 \over m_{\alpha}}, \;
\omega_{B\alpha}={e_{\alpha} B_0 \over m_{\alpha} c}$.
In the cylindrical coordinates $(r,\; \phi,\; z)$ with ${\partial \over
\partial \phi}=0,\; {\partial \over
\partial z}=\,ik $ we have from (\ref{maxth}),(\ref{ref45})

\begin{equation}
\label{ref48}
-ikB_{\phi}\,=\,{4\pi \over c}\,j_r\,-\,{i \omega \over c} E_r,
\end{equation}
$$ ikB_r\,-{dB_r \over dr}\,=\,{4\pi \over c}\,j_{\phi}\,
-\,{i \omega \over c} E_{\phi}, \quad
{1\over r}{d \over dr}(rB_{\phi})\,=\,{4\pi \over c}\,j_z\,-
\,{i \omega \over c} E_z,  $$
\begin{equation}
\label{ref49}
-ikE_{\phi}\,=\,{i\omega \over c} B_r,\,\,\,
ikE_r-{dE_z \over dr}\,=\,{i\omega \over c} B_{\phi},\,\,\,
{1 \over r}{d \over dr} (rE_{\phi})\,=\,{i\omega \over c} B_z.
\end{equation}
We are interested in long wave oscillations with $\omega \ll \omega_{pe}$,
$\omega_{pp}$, $\omega_{Be}$, $\omega_{Bp} $, so approximately

\begin{equation}
\label{ref50}
 j_r\,=\,-{i\over 4 \pi}{\omega_{pp}^2 \omega \over \omega_{Bp}^2} E_r,\,\,\,
j_{\phi}\,=\,-{i\over 4 \pi}{\omega_{pp}^2 \omega \over \omega_{Bp}^2}
 E_{\phi},\,\,\,
j_z\,=\,-{i\over 4 \pi}{\omega_{pe}^2 \over \omega} E_z.
\end{equation}
Substituting (\ref{ref50}) into (\ref{ref48}),(\ref{ref49})
we can see that two types of long waves,
corresponding to different polarizations exist independently: electric-type (E)
waves with $B_z\,=\,0 $ and nonzero $(E_r,\,E_z,\,B_{\phi})$, and
magnetic-type (B) waves with $E_z\,=\,0 $ and nonzero $(B_r,\,B_z,\,E_{\phi})$.
The equations for $E-$ wave has a form

\begin{equation}
\label{ref51}
{d \over dr} \biggl[{1 \over r}{d \over dr} \bigl(rZ_E \bigr) \biggr]\,+\,
\biggl({\omega_{pe}^2 \over \omega^2} - 1 \biggr)\biggl({k^2c_A^2 \over c^2}\,-
\,{\omega^2 \over c^2 }\biggr) Z_E \,=\,0,
\end{equation}

\begin{equation}
\label{ref52}
{1 \over r}{d \over dr} \biggl(r{dE_z \over dr} \biggr)\,+\,
\biggl({\omega_{pe}^2 \over \omega^2} - 1 \biggr)\biggl({k^2c_A^2 \over c^2}\,-
\,{\omega^2 \over c^2} \biggr) E_z \,=\,0,
\end{equation}
where $Z_E\,=\,E_r$ or $B_{\phi}$, and

\begin{equation}
\label{ref53}
c_A^2\,=\,c^2 \left(1\,+\,{\omega_{pp}^2 \over \omega_{Bp}^2}\right)^{-1}\,=\,
c^2\left(1\,+\,{4 \pi \rho c^2 \over B^2}\right)^{-1}
\end{equation}
is a speed of the Alfven waves. The equations for a $B-$ wave has a form

\begin{equation}
\label{ref54}
{d \over dr} \biggl[{1 \over r}{d \over dr} \bigl(rZ_B \bigr) \biggr]\,+\,
\biggl({\omega^2 \over c_A^2} - k^2 \biggr)\,Z_B \,=\,0,
\end{equation}
\begin{equation}
\label{ref55}
{1 \over r}{d \over dr} \biggl(r{dB_z \over dr} \biggr)\,+\,
\biggl({\omega^2 \over c_A^2} - k^2 \biggr)\,B_z \,=\,0,
\end{equation}
where $Z_B\,=\,B_r$ or $E_{\phi}$,
Outside the cylinder equations (\ref{ref51}),(\ref{ref52}),
(\ref{ref54}),(\ref{ref55}) retain their form after
substituting $c$ everywhere instead of $c_A$. Solutions of these equations
are given by Bessel functions:

\begin{equation}
\label{ref56}
Z_{B,E}=Z_{B0,E0}J_1(\kappa_{B,E} r),\;B_Z=B_{Z0}J_0(\kappa_B r),\:
E_Z=E_{Z0}J_0(\kappa_E r),
\end{equation}
where
$\kappa_B^2={\omega^2 \over c_A^2}-k^2, \;\,
  \kappa_E^2=\biggl({\omega_{pe}^2 \over \omega^2}-1 \biggr) \biggl({k^2c_A^2
  \over c^2}-{\omega^2 \over c^2} \biggr)$
inside the cylinder. Outside the same solutions (\ref{ref56}) are valid with
$ \kappa^2={\omega^2 \over c^2}-k^2$,
instead of $\kappa_E^2$ or $\kappa_B^2$.
Discreet values of $\kappa_B$ and $\kappa_E$ are determined by
dispersion equation, obtained from boundary conditions
on the surface of the cylinder.
For $B-$ wave the components $B_r$ and $E_{\phi}$ are continuous on the
boundary and $B_z\,=\,0$ is taken at the boundary $r=r_0-0$. That leads to
the relations

\begin{equation}
\label{ref58}
\kappa_{Bn}r_0\,=\,\lambda_{0n}\;,\;\;\omega_n^2=c_A^2\biggl(k^2+
{\lambda_{0n}^2 \over r_0^2}\biggr).
\end{equation}
For $E-$ wave $E_z$ is continuous and $E_r\,=\,0$ at
$r\,=\,r_0-0$, so we have

\begin{equation}
\label{ref59}
\kappa_{Ei}r_0\,=\,\lambda_{1i},\;\;\;\omega_i^2\,=\,c_A^2 k^2\biggl(1+{
\lambda_{1i}^2c^2 \over r_0^2\omega_{pe}^2} \biggr) ^{-1},
\end{equation}
where $\lambda_{0n}$ and $\lambda_{1i}$ are zero's of the Bessel functions
$J_o(x)$ and $J_1(x)$. We have then for the wave vectors outside the cylinder

\begin{equation}
\label{ref60}
\kappa_n^2={c_A^2 \over c^2}{\lambda_{0n}^2 \over r_0^2 }\, -\, k^2
\biggl(1\,-\,{c_A^2 \over c^2 }\biggr) \,\,\, (B),\quad
\kappa_i^2\,=\,-k^2\biggl(1-{c_A^2/c^2 \over 1+{\lambda_{1i}^2c^2 \over
r_0^2 \omega_{pe}^2}} \biggr) \,\,\, (E).
\end{equation}
It is clear from (\ref{ref53}) that $\kappa_i^2\,<\,0 $ for $E-$ waves, but for each
$k$ there is $n_{\ast}$ such that $\kappa_n^2 \,>\,0$ for $n\,>
\,n_{\ast} $ and Alfven $B-$ wave is transforming into electromagnetic one.
The vacuum values of $B_r(r_0+0)$ and $E_{\phi}(r_0+0)$
at the boundary are obtained from the continuity conditions,
and the outer value of $B_z(r_0+0)$ is found from the last equation in (\ref{ref49}).
The
jump $\Delta B_z\,=\,B_z(r_0+0)$ on the boundary determines the induced
surface
current $i_{\phi}$.
Very long electromagnetic wave emitted with the amplitude $E_{\phi}$  could
accelerate particles up to energies  $\epsilon \, \approx e\,E\,r_0\,\approx
\,\alpha 10^3\,$ ergs for $r_0\,=\,1\,$ pc and $E\,=\,\alpha \cdot 10^{-6} $
in CGS. Real energy of the accelerated particles could be much less due to
radiation losses.

\section{Conclusion}

Similarity of the acceleration mechanisms of the particles in the pulsars and
relativistic jets could be a reason of the similarity of the high energy
radiation around $100$ Mev which have been observed by EGRET in number
of radiopulsars \cite{tho92};
quasars and AGN \cite{tho93}, but in no other compact objects.

\acknowledgements
Author is grateful to the organizers
for support and hospitality.

\end{article}

\end{document}